\documentclass{natureprintstylev4}

\usepackage{astjnlabbrev-nature}

\usepackage{amssymb}
\usepackage{mathptmx}
\usepackage{amsmath}	
\usepackage{gensymb}
\usepackage{color}
\usepackage{enumitem}

\usepackage{epsfig}
\usepackage{color}
\usepackage{graphicx}
\usepackage{longtable}
\usepackage{hyperref}

\usepackage[switch]{lineno}

\usepackage[labelsep=endash]{caption}

\newcommand{\citep}{\cite}
\newcommand{\citet}{\cite}

\def\msun{\,{\rm M}_\odot}

\newcommand{\za}{$\rm \langle z_A \rangle$}
\newcommand{\zb}{$\rm \langle z_B \rangle$}

\newcommand{\dzab}{$\rm \Delta{z_{AB}} \ $}

\title{Possible observational evidence that cosmic filaments spin}

\author{Peng~Wang$^{1*}$, Noam~I.~Libeskind$^{1,2*}$, Elmo~Tempel$^{3}$, Xi~Kang$^{4,5}$, Quan~Guo$^{6}$}

\begin{document}

\maketitle

\begin{affiliations}
\item Leibniz-Institut f\"ur Astrophysik Potsdam, An der Sternwarte 16, D-14482 Potsdam, Germany.
\item University of Lyon; UCB Lyon 1/CNRS/IN2P3; IPN Lyon (IPNL), France.
\item Tartu Observatory, University of Tartu, Observatooriumi 1, 61602 T\~oravere, Estonia.
\item Zhejiang University-Purple Mountain Observatory Joint Research Center for Astronomy, Zhejiang University, Hangzhou 310027, China.
\item Purple Mountain Observatory, No. 10 Yuan Hua Road, 210034 Nanjing, China.
\item Shanghai Astronomical Observatory, Nandan Road 80, Shanghai 200030, China.
\end{affiliations}
\\
 
\begin{abstract}  %
Most cosmological structures in the universe spin. Although structures in the universe form on a wide variety of scales from small dwarf galaxies to large super clusters, the generation of angular momentum across these scales is poorly understood. We have investigated the possibility that filaments of galaxies - cylindrical tendrils of matter hundreds of millions of light-years across, are themselves spinning. By stacking thousands of filaments together and examining the velocity of galaxies perpendicular to the filament's axis (via their red and blue shift), we have found that these objects too display motion consistent with rotation making them the largest objects known to have angular momentum. The strength of the rotation signal is directly dependent on the viewing angle and the dynamical state of the filament. Just as it is easiest to measure rotation in a spinning disk galaxy viewed edge on, so too is filament rotation clearly detected under similar geometric alignment. Furthermore, the mass of the haloes that sit at either end of the filaments also increases the spin speed. The more massive the haloes, the more rotation is detected. These results signify that angular momentum can be generated on unprecedented scales.
\end{abstract}
 
\section*{Introduction}
How angular momentum is generated in a cosmological context is one of the key unsolved problems of cosmology. In the standard model of structure formation, small overdensities present in the early universe grow via gravitational instability as matter flows from under to overdense regions. Such a potential flow is irrotational or curl-free: there is no primordial rotation in the early universe and angular momentum must be generated as structures form.

Tidal torque theory \citep{1969ApJ...155..393P, 1970Afz.....6..581D, 1970Ap......6..320D, 1984ApJ...286...38W,1987ApJ...319..575B, 1993ApJ...418..544V, 1996ApJS..103....1B, 1999A&A...343..663P, 2002MNRAS.332..325P} provides one explanation - the misalignment of the inertia tensor of a gravitationally collapsing region of space with the tidal (shear) field can give rise to torques which spin up the collapsing material \citep{1969ApJ...155..393P, 1970Ap......6..320D, 1999A&A...343..663P}. Such an explanation is valid only in the linear regime namely in limit where density perturbations are small with respect to the mean and where flows are laminar. As a collapsing region reaches turn-around tidal torques cease to be effective and the final angular momentum of a collapsed region is far from what tidal torque theory would predict \citep{2002MNRAS.332..325P, 2016IAUS..308..421P}. Although one recent study \citep{2020arXiv200304800M} has demonstrated that galaxy spin direction (i.e. clockwise versus counter clockwise) can be predicted from initial conditions, revealing a critical clue to the non-linear acquisition of angular momentum, our understanding of spin magnitude, direction and history remains in its infancy. Regions that are still in the linear or quasi-linear phase of collapse could provide a better stage for the application of tidal torque theory. 

Cosmic filaments \citep{1996Natur.380..603B}, being quasi-linear extended topographical features of the galaxy distribution, provide such an environment. Yet, owing to the challenges in characterizing and identifying such objects, potential rotation on the scales of cosmic filaments has been discussed \citep{2020OJAp....3E...3N} but never measured until now. 

It is known that the cosmic web in general and filaments, in particular, are intimately connected with galaxy formation and evolution \citep{2012MNRAS.427.3320C, 2019OJAp....2E...7A}. They also have a strong effect on galaxy spin \citep{2000ApJ...532L...5L, 2007ApJ...671.1248L, 2007ApJ...655L...5A, 2013MNRAS.428.2489L, 2013MNRAS.428.1827T, 2015ApJ...798...17Z}, often regulating the direction of how galaxies \cite{2013ApJ...775L..42T, 2014MNRAS.444.1453D, 2014MNRAS.443.1090F, 2015MNRAS.452.3369C, 2016MNRAS.457..695P, 2018ApJ...866..138W, 2018MNRAS.481.4753C, 2019ApJ...876...52K, 2020MNRAS.491.2864W, 2020MNRAS.493..362K} and their dark matter halos rotate \cite{2007MNRAS.375..489H, 2007MNRAS.381...41H, 2009ApJ...706..747Z, 2013ApJ...762...72T, 2015MNRAS.446.2744L,  2017MNRAS.468L.123W, 2018MNRAS.473.1562W, 2019ApJ...872...37L, 2018MNRAS.481..414G, 2019MNRAS.487.1607G}. However, it is not known whether the current understanding of structure formation predicts that filaments themselves, being uncollapsed quasi-linear objects should spin. A recent study (published on the ArXiv while this draft was being finalized) \citep{2020arXiv200602418X} examined the velocity field around galactic filaments defined by halo pairs in a large $N$-body simulation and found a statistically significant rotation signal. This is an intriguing finding and, although filaments and their rotation speed are defined differently, the current work is in which the observed galaxy distribution is examined in a bid to find possible filament rotation was  partly motivated by the theoretical suggestion that filaments may spin \citep{2020arXiv200602418X}.

\section*{Results}
\begin{figure*}[!ht]
\centerline{\includegraphics[width=0.9\textwidth, trim=0.0cm 0.0cm 0.0cm 0.0cm, clip=true]{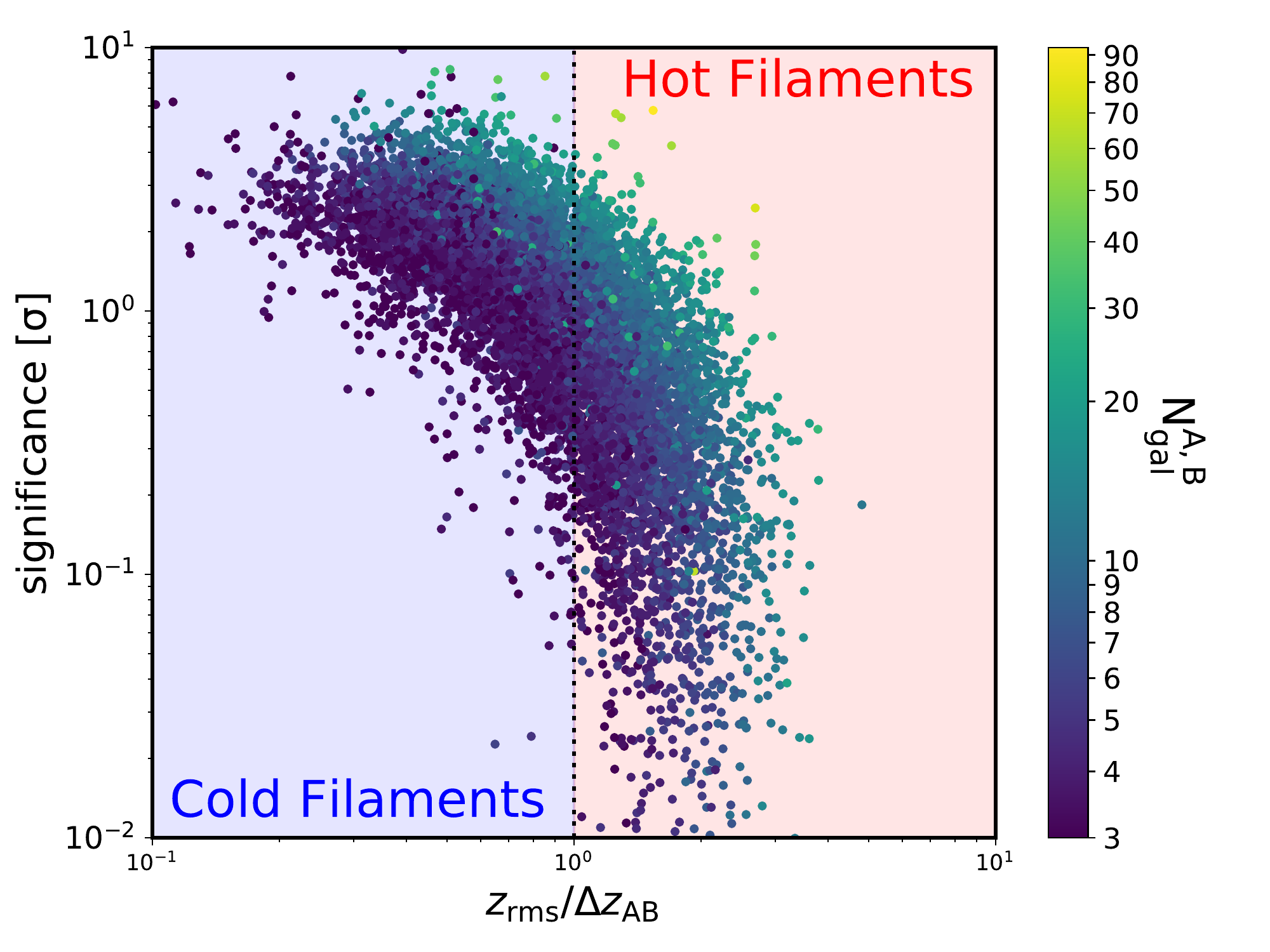}}
\caption{The statistical significance of \dzab  being consistent with random is shown as a function of {the filament dynamical ``temperature'',} $z_{\rm rms}/\Delta z_{\rm AB}$, {in which \dzab is the redshift difference of galaxies between the approaching and receding regions of each filament.} The higher this quantity, the more unlikely it is that \dzab is a random occurrence. Each filament is colored codes by the number of galaxies it contains, with yellow being rich filaments and purple being poor filaments as designated by the color bar. At a given value of $z_{\rm rms}/\Delta z_{\rm AB}$, the \dzab seen in richer filaments is more statistically inconsistent with random. `Cold' or `hot' filaments are separated by $z_{\rm rms}/\Delta z_{\rm AB}=1$.} 
\label{fig:sig_zab}
\end{figure*}

After segmenting the galaxy distribution into filaments using a marked point process known as the Bisous model\cite{2007JRSSC..56....1S}, each filament can be approximated by rectangle on the sky and thus the galaxies within it may be divided into two regions on either side of the filament spine. The mean redshift difference \dzab of galaxies between two regions are considered as a proxy for the line of sight velocity difference and hence for the filament spin signal. Any measured value of \dzab needs to be assigned a significance based on a randomization procedure (explained in the methods section). Figure~\ref{fig:sig_zab} shows the statistical significance of the measureed \dzab as a function of  $z_{\rm rms}/\Delta z_{\rm AB}$ - a proxy for the dynamical ``temperature'' of the filament (see method section for more details).
The number of galaxies in each region is denoted by color. Two salient points can be gleaned here. First, The more galaxies in a given filament, the more inconsistent the redshift difference \dzab is with random. Second, (as expected) the colder the filament, the more inconsistent the redshift difference is with random. This second point is a generalization - cold filaments with $z_{\rm rms}/\Delta z_{\rm AB} < 1$ and few galaxies, can have redshift differences only weakly inconsistent with random expectations. However one may note that as a trend, the colder the filament, the more significant the redshift difference is. In other words if \dzab is considered a proxy for filament spin, we observe a spectrum of filaments from dynamically hot that are consistent with random to dynamically cold filaments that are completely inconsistent with random at the many sigma level. Note that even for dynamically hot filaments, there are a few that are highly inconsistent with random. The reader will note that the wide distribution of significance seen at a given value of $z_{\rm rms}/\Delta z_{\rm AB}$, is also a reflection of the distribution of inclination angle made by the filament axis with the line of sight (see figure A1).

The cumulative distribution of \dzab for both the entire observed filament sample as well as the randomized trials, is shown in the three panels of Figure~\ref{fig:delta_zab}. The cumulative distribution is shown for all filaments (left), filaments whose axis is inclined by $\cos\phi<0.5$ to the line of sight (middle) and dynamically cold filaments ($z_{\rm rms}/\Delta z_{\rm AB} < 1$) inclined by $\cos\phi<0.2$ (right). 
The reader will note that even when examining all filaments where inclination angle is completely ignored (and hence includes filaments viewed along their axis which will likely weaken the signal), the full distribution of \dzab is inconsistent with randomization tests. This inconsistency increases when considering filaments with $\cos\phi<0.5$ and  $\cos\phi<0.2$ that are cold. To quantify the  statistical significance of the cumulative distribution one may simply measure, in units of the 10,000 random trial's standard deviation, how far the measured signal is from the mean randomized signal. This is plotted in the upper panels of Figure~\ref{fig:delta_zab} and shows that randomized trails are statistically inconsistent with the measured signal at very high confidence. In other words, shuffling the redshifts of galaxies in a filament is unlikely to produce redshift differences as great as that observed.

\begin{figure*}[!ht]
\includegraphics[width=0.9\textwidth, trim=0.0cm 0.0cm 0.0cm 0.0cm, clip=true]{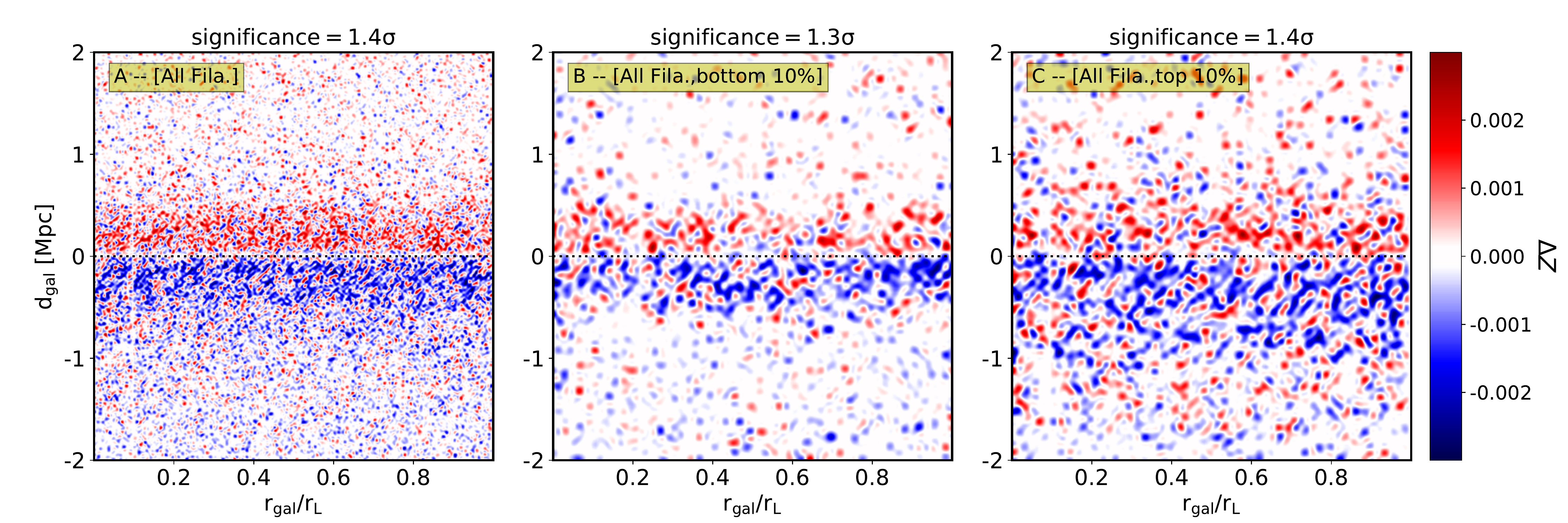}
\includegraphics[width=0.9\textwidth, trim=0.0cm 0.0cm 0.0cm 0.0cm clip=true]{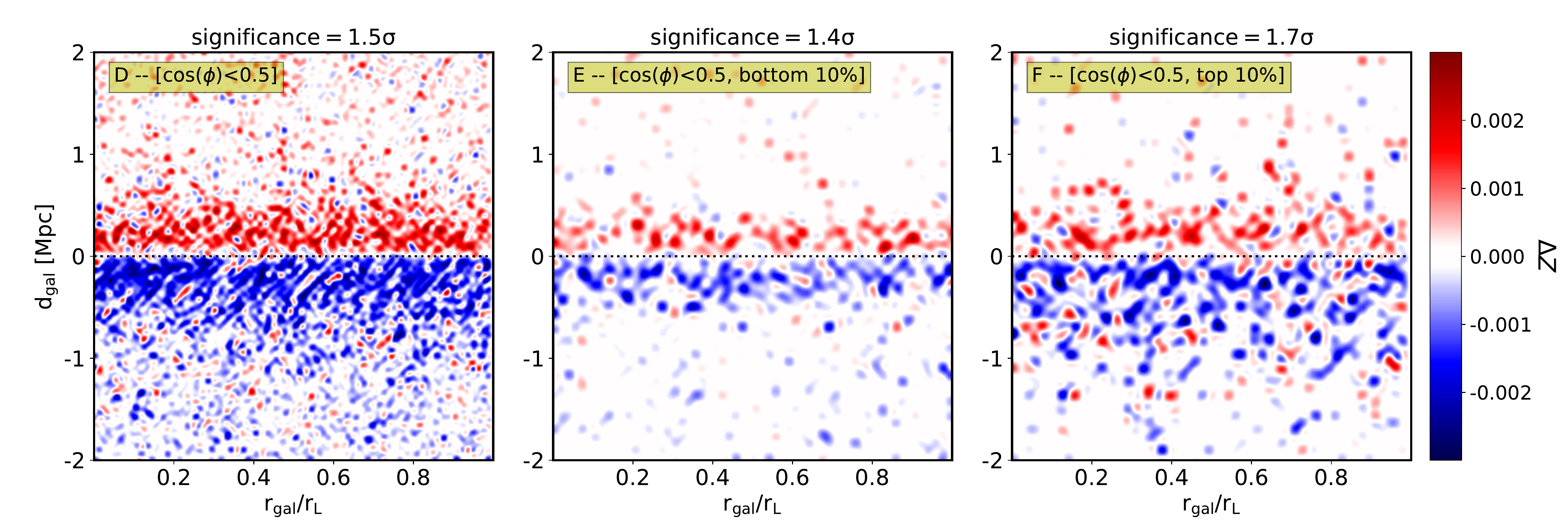}
\includegraphics[width=0.9\textwidth, trim=0.0cm 0.0cm 0.0cm 0.0cm, clip=true]{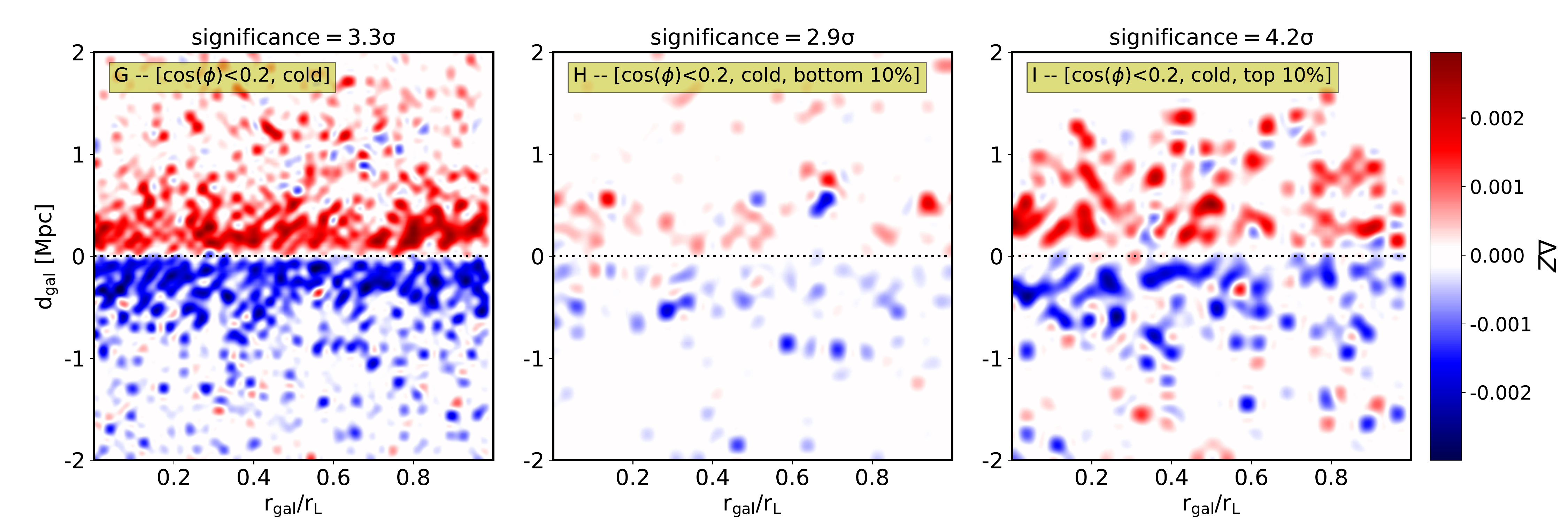}
\caption{The stacked rotation signal of filaments. Galaxies' position $r_{\rm gal}$ long the filament are scaled by the length $r_{\rm L}$ of the filament. $d_{\rm gal}$ is the distance of galaxies to the filament axis. Region A (defined as the region with greater mean redshift) is shown in the upper part of each plot, while region B is shown in the lower part. From top to bottom, each row shown the stacked rotation signal for all filaments, filaments whose spine subtend an angle $cos(\phi)<0.5$ with the line of sight and filament with $cos(\phi)<0.2$ and which have $z_{\rm rms}$/$\Delta z_{\rm AB}<1$. These three samples of filaments are subdivided according the group mass they are point to. Middle and right columns show the stacked rotation signal for filaments with the smallest and largest 10 percentile filament end point mass. The redshift difference is coded in color bar.}
\label{fig:stackedsignal}
\end{figure*}

The redshift difference (a proxy for the rotation signal) stacked across various (sub-) samples, is presented in Figure~\ref{fig:stackedsignal}. The following conventions have been adopted. Region A (defined as the region with greater mean redshift) is plotted in the upper part of each plot, while region B is plotted in the lower part. The position of each galaxy along and perpendicular to the filament axis is shown on the x and y axis, respectively and, along the x axis, is normalized to the filament's length. Each galaxy is colored by its redshift difference $\Delta z$, with respect to the mean redshift of all galaxies in the filament according to the color bar at right. In the ideal situation where all galaxies exhibit circular or helical motion about the filament axis, such a plot would only have red points in  the upper part and blue points in the lower part. The statistical significance of each (sub-) sample is denoted by the significance, in units of $\sigma$ and indicated on top of each panel.

In Figure~\ref{fig:stackedsignal} left column (i.e. panels, A, D, G) we show the stacked rotation signal for all filaments, filaments whose axis subtends an angle $\cos\phi<0.5$ with the line of sight and filaments whose axis subtends an angle $\cos\phi<0.2$ with the line of sight and which have $z_{\rm rms}/\Delta z_{\rm AB} <1$.  Comparing panel (A) with (D) the reader will note what has been mentioned before namely that merely changing the inclination angle increases the signal. Panel (G) shows a very strong rotation signal - at 3.3$\sigma$ - when considering dynamically cold filaments that are mostly perpendicular to the line of sight the rotation signal becomes very convincing. 

Since filaments are long tendrils of galaxies often connecting nodes of the cosmic web, the mass of the two halos closest to the filament's two end points are examined for a possible correlation with the signal strength we measure. We thus perform the following procedure. We search the group catalogue \citep{2017A&A...602A.100T} for the two groups that are closest to the filament's two end points. The reader is referred to that paper for details. In sum, the galaxy groups are detected with a modified Friend-of-Friend (FoF) algorithm, where the FoF groups are divided into subgroups so that each subgroup can be considered as a virialized structure. The mass of each group is estimated using the virial theorem, the size of the group in the sky and the group velocity dispersion along the line of sight. By summing these two group masses we can ascribe a single mass abutting each filament. We call this mass the ``filament end point mass''.

The three samples of filaments can then be subdivided according to this characteristic. Figure~\ref{fig:stackedsignal} middle and right columns show the stacked signal for the 10 percent of the filaments with the smallest filament end point mass (ie panels B,E,H) and largest 10 percent filament end point mass (panels C, F, I) associated halo end points, respectively. The limits of the 10th and 90th percentile group masses of panel H and I are $\sim10^{10.4}$ and $\sim10^{14.3}$, in units of $\msun$, respectively. 

A clear dependence on filament end point mass is seen for all three samples in Figure~\ref{fig:stackedsignal}. The smaller the filament end point mass, the weaker the rotation signal. The filaments with the largest end point mass, most perpendicular to the line of sight, and dynamically cold, show an outstanding 4.2$\sigma$ inconsistency with the null hypothesis of random origin. The fact that the mass of the clusters at the end of each filament directly affects the measured signal, is evidence that the signal we are measuring is a physical effect related to gravitational dynamics.

\begin{figure*}[!ht]
\centerline{
\includegraphics[width=0.9\textwidth, trim=0.0cm 0.0cm 0.0cm 0.0cm, clip=true]{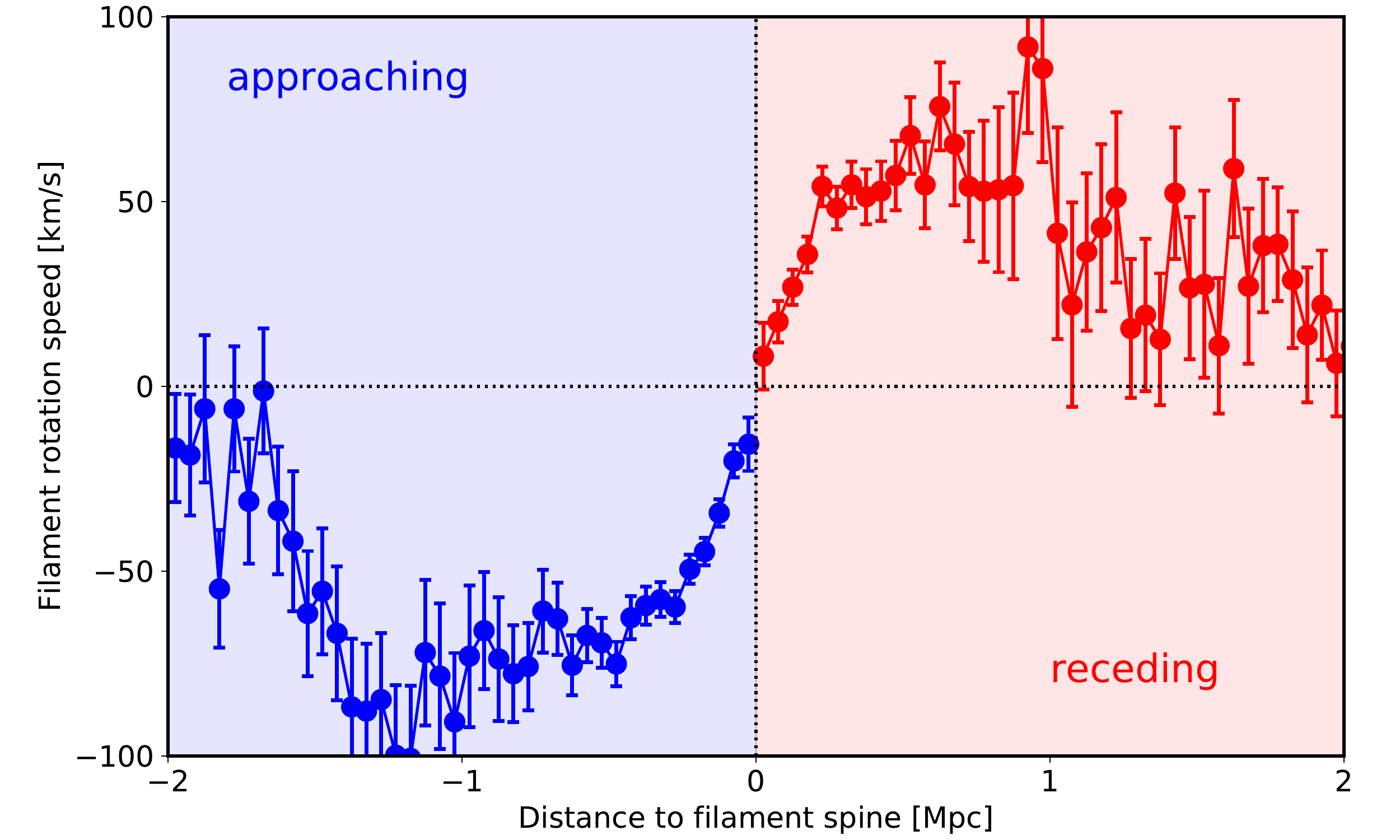}
}
\caption{The rotation curve of filaments as a function of the distance between galaxies to filament spine. 
The rotation speed is calculate by $c\times \Delta {z_{i}}$, where $\Delta {z_{i}}$ is the redshift difference of galaxies at given distance with respect to the mean redshift of all galaxies in the filament. The distance of galaxies in the region A (or receding region, defined as the region with greater mean redshift) is marked as positive values, while the distance of galaxies in the region B (or approaching region, defined as the region with smaller mean redshift) is marked as negative values. Error bars represent the standard deviation about the mean.}
\label{fig:rotation_curve}
\end{figure*}

A ``rotation curve'' can be constructed for the stacked filament sample by converting the redshift difference into a velocity difference computed as $c\times \Delta {z_{i}}$, where $\Delta {z_{i}}$ is the redshift difference between all the galaxies at given distance and the mean redshift of all galaxies in the filament. This is shown in Fig.~\ref{fig:rotation_curve}.  We adopt the typical convention that negative (positive) velocities corresponds to approaching (receding) motion. Fig.~\ref{fig:rotation_curve} shows that rotation speed increases with distance, reaching a peak of around $\sim 100km/s$ at a distance $\sim1$ Mpc. It then decreases with distance, tending to zero at a distance of greater than $\sim$ 2 Mpc. Incidentally this behavior motivates the choice of filament thickness in this study, and should be kept in mind in future studies of the dynamics of cosmic filaments.

\section*{Summary \& Discussion}
\begin{figure*}[!ht]
\centerline{
\includegraphics[width=0.32\textwidth, trim=0.0cm 0.0cm 0.0cm 0.0cm, clip=true]{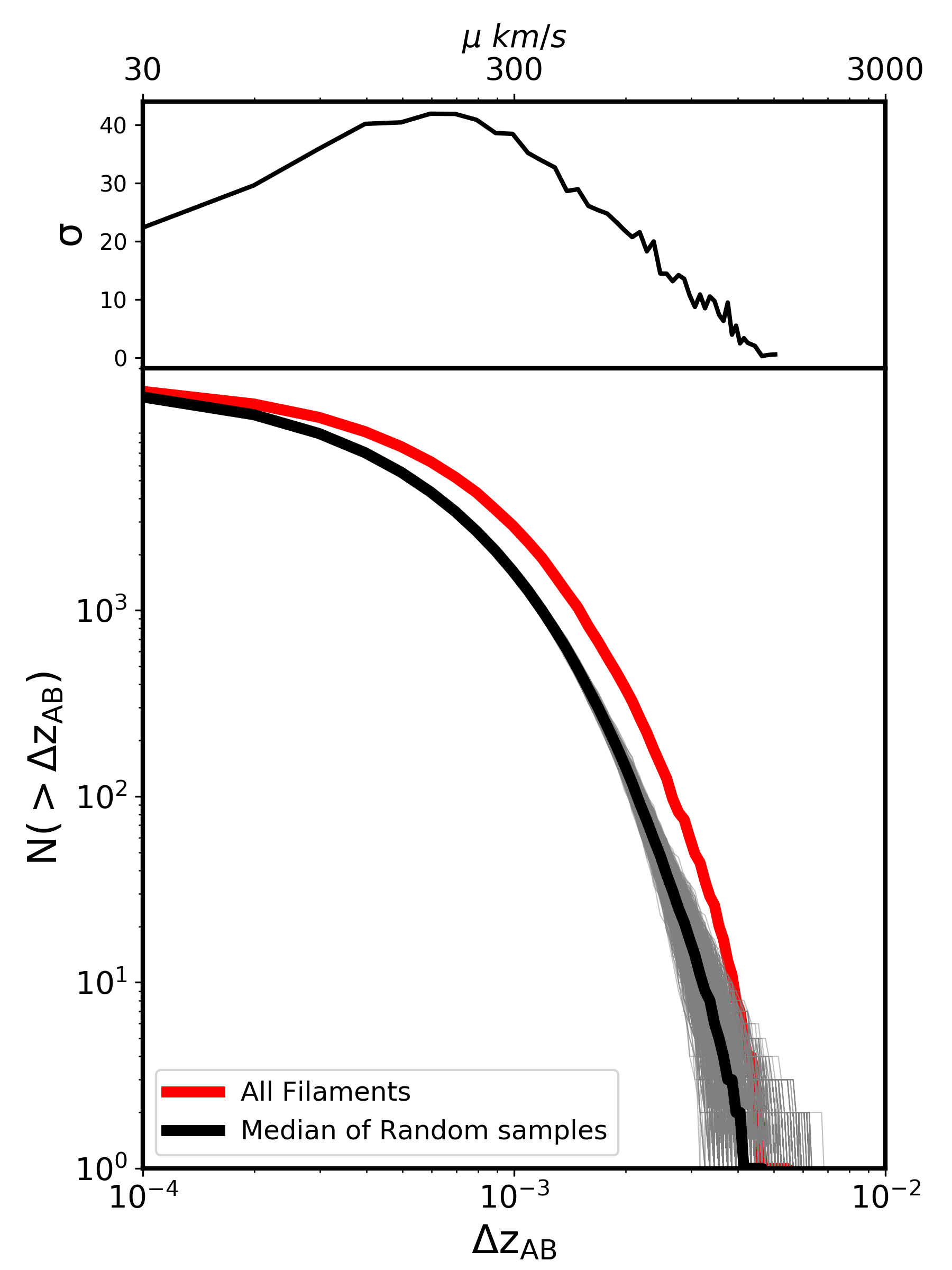}
\includegraphics[width=0.32\textwidth, trim=0.0cm 0.0cm 0.0cm 0.0cm, clip=true]{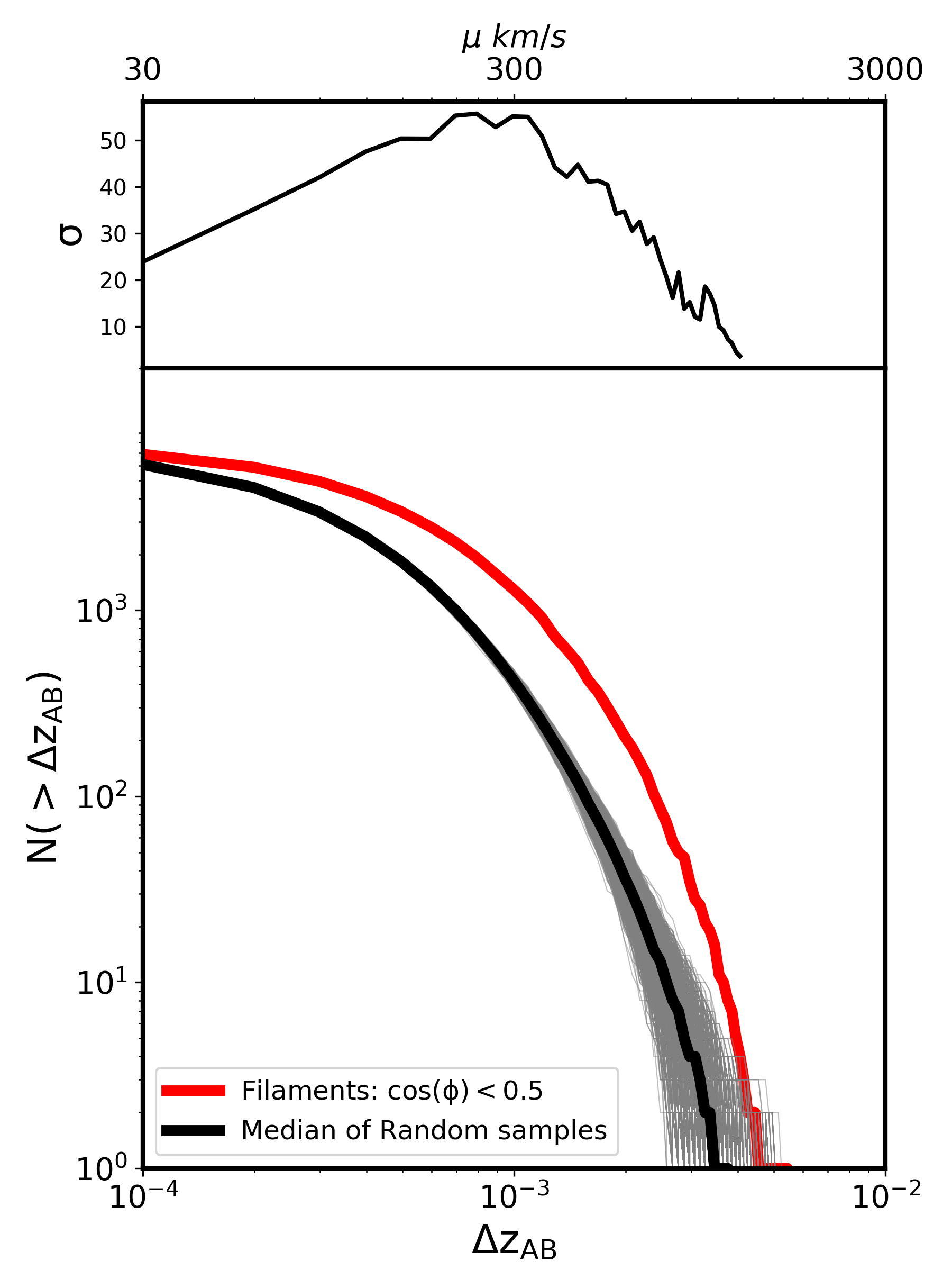}
\includegraphics[width=0.32\textwidth, trim=0.0cm 0.0cm 0.0cm 0.0cm, clip=true]{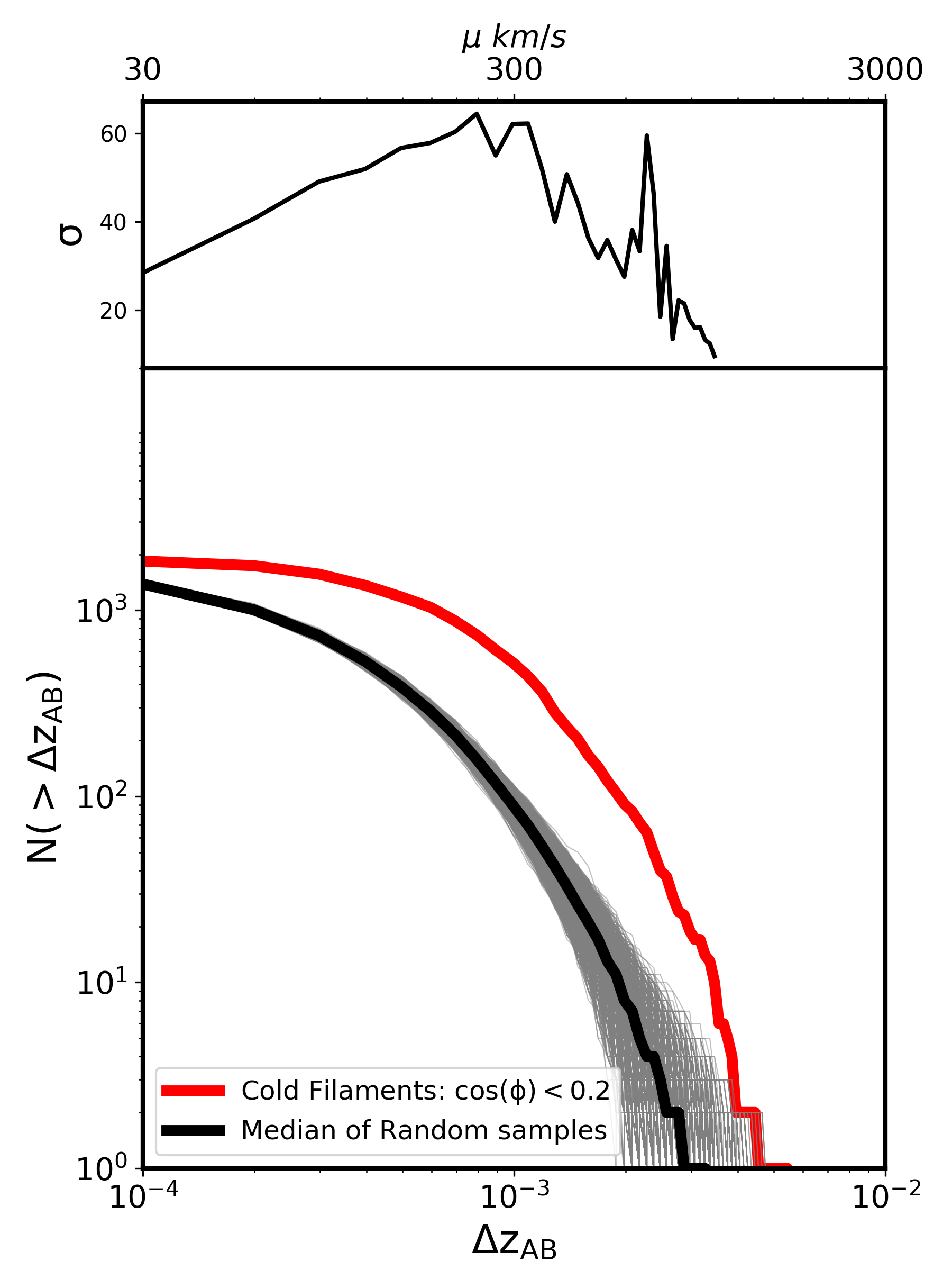}
}
\caption{The cumulative distribution of \dzab, the redshift difference {of galaxies in} the approaching and receding regions of each filament. Left panel: all filaments irrespective of viewing angle. Middle panel: filaments whose axis subtends an angle $\cos\phi<0.2$ with the line of sight. Right panel: filaments whose axis subtends an angle $\cos\phi<0.5$ with the line of sight and which are dynamically cold, namely $z_{\rm rms}/\Delta z_{\rm AB} <1$. The red solid line shows the distribution of observed filaments, while the 10,000 grey lines indicate the expected distribution from randomized redshifts. The median value of these 10,000 random samples is shown as the black solid line. The upper panels measure, as a function of \dzab , the distance, in units of the standard deviation of the randomized distribution, between the measured curve and the mean of the random distributions. The upper x-axis displays the rotation velocity of the filament in km/s calculated as $\mu=c\times\rm \Delta{z_{AB}}$.}
\label{fig:delta_zab}
\end{figure*}

The results presented in this paper are consistent with the detection of a signal one would expect if filaments rotated. What is measured and presented here is the redshift difference between two regions on either side of a hypothesized spin axis that is coincident with the filament spine. The full distribution of this quantity (Figure~\ref{fig:delta_zab}) is inconsistent with random regardless of the viewing angle formed with the line of sight but does strengthen as more perpendicular filaments are examined. It also strengthens when considering dynamically cold filaments - filaments whose galaxies have small rms values of their redshift. We emphasize that this is not a trivial finding. There is no reason -- besides an inherent rotation -- that filaments that are dynamically cold should exhibit such a signal. 

One may draw an analogy with galaxies here: a disk galaxy viewed edge on shows a strong \dzab (consider \dzab as the Doppler shift from stars on the receding and approaching edge of the galaxy). The value (and significance as measured by the likelihood that it is random) of such a \dzab would decrease as the viewing angle of the galaxy is increased until the galaxy is seen face on. At this point the galaxy would appear ``hot'' - the motion of stars along the line of sight being random.  Additionally, there is a degeneracy between morphology and shape in such situations: a very thin galaxy viewed with a large inclination angle appears ellipsoidal. Since not all galaxies are rotationally supported, some galaxies that appear ellipsoidal will be inclined rotating thin disks while others will be dispersion supported ellipsoids with little inherent rotation. It is this combination of inclination angle and inherent rotation which explains the spectrum of $z_{\rm rms}/\Delta z_{\rm AB}$ and significance seen in both the galaxy distribution as well as the spin of filaments. 

 We note that it is the unique combination of small $z_{\rm rms}$ and large $\Delta z_{\rm AB}$ which returns a strong statistically significant signal. In other words there are filaments with large $z_{\rm rms}$ and small $\Delta z_{\rm AB}$ which clearly do not show any signal. This work does not predict that every single filament in the universe is rotating, rather that there are subs-samples - intimately connected to the viewing angle end point mass - which show a clear signal consistent with rotation. This is the main finding of this work. Again, we note that it is not trivial that filaments with low values of  $z_{\rm rms}$ exhibit large values of  $\Delta z_{\rm AB}$. This is only expected if filaments rotate.

 {According to the zel'dovich approximation \citep{1970A&A.....5...84Z} the motion of galaxy’s is accompanied by larger scale matter flows \citep{1991QJRAS..32...85I, 2014MNRAS.441.2923C}. The mass and velocity flows are usually transported in a “hierarchical” way: at first, matter collapses along the principle axis of compression  forming great cosmic walls. Matter then flows in the wall plane along the direction of the intermediate axis of compression to form filaments. The final stage is the full 3 dimensional collapse of an aspherical anisotrpoic density perturbation wherein matter collapses and flows along the filament axes to form clusters. Under such a model of mass flow, one may expect that filament spin is formed at the second stage from the compression along the intermediate axis of compression. As such mass streams in the sheet towards the filament from both sides and could explain the observed signal. In other words the signal we measure is both consistent with a shear caused by galaxy motion in a sheet, or the spin of a filament embedded in that sheet. These two views are not contradictory with each other, since in the picture described above the spin of a filament is engendered by the shear of the sheet in which it is embedded.}

When the sample of filaments was divided into sub-samples based on the mass of the halos that sit at the ends of the filaments, those filaments with the most massive clusters have more significantly spinning signals than either the full sample or the filaments abutting the least massive halos. This gives a hint towards what may be driving the filament spin signal. It is also consistent with the idea that the spin is due to tidal fields {influenced by the presence of massive aspherical density perturbations. The existence of a mass dependency is also strong evidence that what we have observed is not a systematic effect due to (e.g.) cosmological expansion. Any such systematic effects on the measurement of the redshift difference and its association with a velocity difference, would not manifest as a mass dependency.} The mass dependency opens the door to a number of important questions as to the origin of this rotation or shear, questions that will be examined in future (theoretical) work such as: how is the measured effect engendered by the formation of clusters abutting filaments? Is there an angular momentum correlation between spinning filaments and nearby clusters? and perhaps most importantly: at which stage in structure formation  do filaments spin up, and how does this affect the galaxies within them?

We know that from studies of the spatial distributions of smaller galaxies that inhabit the regions between large galaxies, that they are accreted in a two fold process: first onto the filament connecting the two larger halos and then secondly an accretion along the line connecting the halos. In other words galaxies are first accreted from a direction perpendicular to the filament spine \citep{2015ApJ...813....6K, 2018MNRAS.473.1562W}, and then along the spine \citep{2018MNRAS.473.1562W, 2014MNRAS.443.1274L, 2015ApJ...807...37S}. Assuming filaments can be approximated by a cylindrical potential such a flow would naturally give rise to an rotation about the filament axis, followed possibly by helical accretion. This is consistent with the theoretical picture formed by  \cite{2020arXiv200602418X} who's analysis of numerical simulations indicate that the angular momentum of filaments is generated by those particles which collapse along the filament's potential. Taken together, this paper and  \cite{2020arXiv200602418X} demonstrate that angular momentum can be generated on unprecedented scales, opening the door to a new understanding of cosmic spin.

\begin{flushleft}
{\bf \large References}
\end{flushleft}

\begin{addendum}

\item[Author Information] 
Correspondence and requests for materials should be addressed to P.W. (pwang@aip.de) or N.I.L (nlibeskind@aip.de).

\item[Acknowledgements] 
We acknowledge stimulating discussions with Marcel S. Pawlowski, Stefan Gottl\"oeber, Yehuda Hoffman and Zhao-zhou Li. We acknowledge Rain Kipper for a double-blind examination at the beginning of this project. P.W. especially acknowledge his wife Nan Wang and daughter Tian-en Wang for their understanding and their help of his home office during the COVID-19 pandemic. P.W., N.I.L., X.K. and Q.G. acknowledge support from the joint Sino-German DFG research Project ``The Cosmic Web and its impact on galaxy formation and alignment'' (DFG-LI 2015/5-1, NSFC No. 11861131006). N.I.L acknowledges financial support of the Project IDEXLYON at the University of Lyon under the Investments for the Future Program (ANR-16-IDEX-0005). E.T. was supported by ETAg grants IUT40-2, PRG1006 and by EU through the ERDF CoE TK133. X.K. acknowledges financial supported by the NSFC (No. 11825303, 11333008), the 973 program (No. 2015CB857003), also partially supported by China Manned Space Program through its Space Application System. Q.G. acknowledges financial support of  Shanghai Pujiang Program (No.19PJ1410700).

\item[Author Contributions]
P.W. carried out all measurements and quantitative analysis, contributing to the writing of the method and results.
N.I.L. conceived the idea of the project and led the writing.
E.T. contributed observational data, filament catalogue and essential discussions.
X.K. and Q.G. contributed significantly to the overall science interpretation and essential discussions.
All co-authors contributed by their varied contributions to the science interpretation, data analysis
All co-authors contributed to the commenting on this manuscript as part of an internal review process.  

\item[Competing interests statement] The authors declare no competing interests.

\end{addendum}


\clearpage

\methods

The method presented here is fairly {straightforward}. We begin by using a publicly available catalog of filaments constructed from the SDSS DR12 galaxy survey \citep{2011AJ....142...72E, 2015ApJS..219...12A}. Such a catalog is built using the Bisous algorithm \citep{2007JRSSC..56....1S} which identifies curvi-linear structures in the galaxy distribution\footnote{The Bisous algorithm is not astronomy specific and was developed to identify roads in satellite imaging.} \citep{2014MNRAS.438.3465T, 2016A&C....16...17T,2018MNRAS.473.1195L}. Each filament is approximated by a cylinder with a given axis that makes a {viewing} angle $\phi$ with the line of sight. Our hypothesis rests on the assumption that if filaments rotate then there should be a statistically significant component of a galaxy's velocity that is perpendicular to the filament spine. Furthermore, this component should have opposite signs on either side of the filament spine: one receding and one approaching. In order to detect such a signal in principle we would only want to examine filaments whose spine is perpendicular to the line of sight such that a galaxy's redshift (a proxy for the radial component of its velocity) would coincide with its angular velocity. However restricting the filament sample this way dramatically reduces the sample size. Therefore we divide the filament catalog according to viewing angle into three bins: all filaments irrespective of viewing angle, those inclined by more than 60 degrees ($\cos\phi < 0.5$) with respect to the line of sight, those dynamically cold (see blow) and viewing angle more than $\sim$80 degrees ($\cos\phi < 0.2$). This last sample is chosen for the reason that as demonstrated in Figure~\ref{fig:zrms_dzab_angle}, filaments inclined by more than $\sim$80 degrees are perpendicular enough that the measured value of $z_{\rm rms}/\Delta z_{\rm AB}$ is likely a good approximation to the inherent one. A consequence of this is there are more cold filaments with $z_{\rm rms}/\Delta z_{\rm AB} < 1$ than hot filaments with $z_{\rm rms}/\Delta z_{\rm AB}>1$  inclined by $\cos\phi<0.2$.    In the {appendix section} the effect of viewing angle is examined in more detail. {We find that the filaments whose spines are closer to perpendicular to the line of sight have more significant rotation signals (see figure A1).}

Each filament axis delineates two regions on either side of the filament spine. The Bisous filament's have an inherent scale of around 1 Mpc. However all galaxies which are within a {2 Mpc} distance from the filament axis are considered to be in one of these two regions. Galaxies further afield are not considered. We note that most of the galaxies that are within {2 Mpc} are within 1Mpc and that including galaxies further away from the filament spine will ``dilute'' the signal as the regions further from the filament axis will, by construction, include random interlopers.

The mean redshift of galaxies in both regions is calculated. The region with the greater mean redshift is arbitrarily called region A, the region with smaller mean redshift is called region B. The redshift difference  \dzab=\za-\zb ~is computed. The relative speed is then simply $\mu=c$\dzab, where $c$ is the speed of light. Next we examine if the filament is dynamically hot or cold by using the redshift $z$ of all galaxies in each filament to compute the rms namely $z_{\rm rms}$. This is then compared with \dzab. Dynamically hot filaments have $z_{\rm rms}>$\dzab while cold filaments have rms values less than the redshift difference. Such a definition precludes the ability to search for a signal consistent with rotation in hot filaments since it is by definition flawed to ascribe a physical significance to a \dzab if it is less than $z_{\rm rms}$. Thus we do not expect to be able to find a discernible rotation signal in dynamically hot filaments.

The median value of $z_{\rm rms}/\Delta z_{\rm AB}$, (as well the standard deviation) is shown in Figure~\ref{fig:zrms_dzab_angle} as a function of the angle subtended between the filament spine and the line of sight. Two important points can be observed by this plot: (1) the filament's dynamical temperature ($z_{\rm rms}/\Delta z_{\rm AB}$) increases as filaments become parallel to the line of sight. Because the actual value of $z_{\rm rms}/\Delta z_{\rm AB}$ for a given filament is independent of $\phi$, this fact demonstrates that the viewing angle is critical in measuring this quantity and that $z_{\rm rms}/\Delta z_{\rm AB}$ can indeed be used to measure rotation  for filaments that are close to perpendicular to the line of sight. Furthermore such a relationship is natural if a significant component of a galaxy's velocity is perpendicular to the filament spine and consistent with rotation or shear. (2) The standard deviation of $z_{\rm rms}/\Delta z_{\rm AB}$ increases as filaments become parallel to the line of sight. This expected if in addition to rotational motion about the filament spine, galaxies are also traveling along the filament axis - perhaps with helical motion.

A lower limit on the number of galaxies in each region of A and B of each filament must be assumed. Without such a limit we would introduce pathological cases into the analysis, namely filaments with no or just one or two galaxies in region A or B. Therefore we only consider filaments with at least 3 galaxies in each region - it is difficult to justify the computation of an rms for anything less than this. The three galaxy per region A and B is a conservative choice and is the least arbitrary limit that can be applied. Despite this choice, the signal is examined as a function of galaxy number in the {appendix section}. In sum: increasing the minimum number of galaxies in each region decreases the sample size and increases the significance of the measured signal.The final fiducial sample consists {of 17,181} filaments, and {213,625} galaxies in total.

A statistical significance to each measured value of \dzab \ must be obtained to ensure that the measured signal is not simply the result of stochasticity. In other words we wish to ask the question ``What is the chance that the observed distribution or \dzab could be reproduced given a random distribution of redshifts (in a given filament)?'' In order to answer this question, the following test is performed. The redshifts of all galaxies in a given filament are randomly shuffled, keeping the position fixed. This is performed 10,000 times per filament.\footnote{In the poorest filaments with only 6 galaxies (say), there are less than 10,000 unique permutations. However performing this test 10,000 times, although redundant, will not affect the determination of the significance.} For each of these trials, the procedure described above is carried out: namely two regions are defined, their mean redshifts are measured and $\Delta{\rm z}^{\rm random}_{\rm AB}$ is computed. The ``real'' \dzab can then be compared to the distribution of random $\Delta{\rm z}^{\rm random}_{\rm AB}$ on a filament by filament basis. The chance that a measured \dzab is consistent with random can then be quantified by where (how many standard deviations from the mean) the measured \dzab is, in the distribution of $\Delta{\rm z}^{\rm random}_{\rm AB}$.

\appendix
\renewcommand\thefigure{\thesection\arabic{figure}}   
\setcounter{figure}{0} 

\section{The effect of filament viewing angle and galaxy number}\label{sec:appendix}
In this section we examine how the signal changes based on two arbitrary choices we are forced to take in this work:  the orientation of filaments with respect to the line of sight and the lower limit for the number of galaxies in each region to be considered.

\begin{figure*}[!ht]
 \centerline{\includegraphics[width=0.9\textwidth, trim=0.25cm 0.25cm 0.25cm 0.5cm,clip=true]{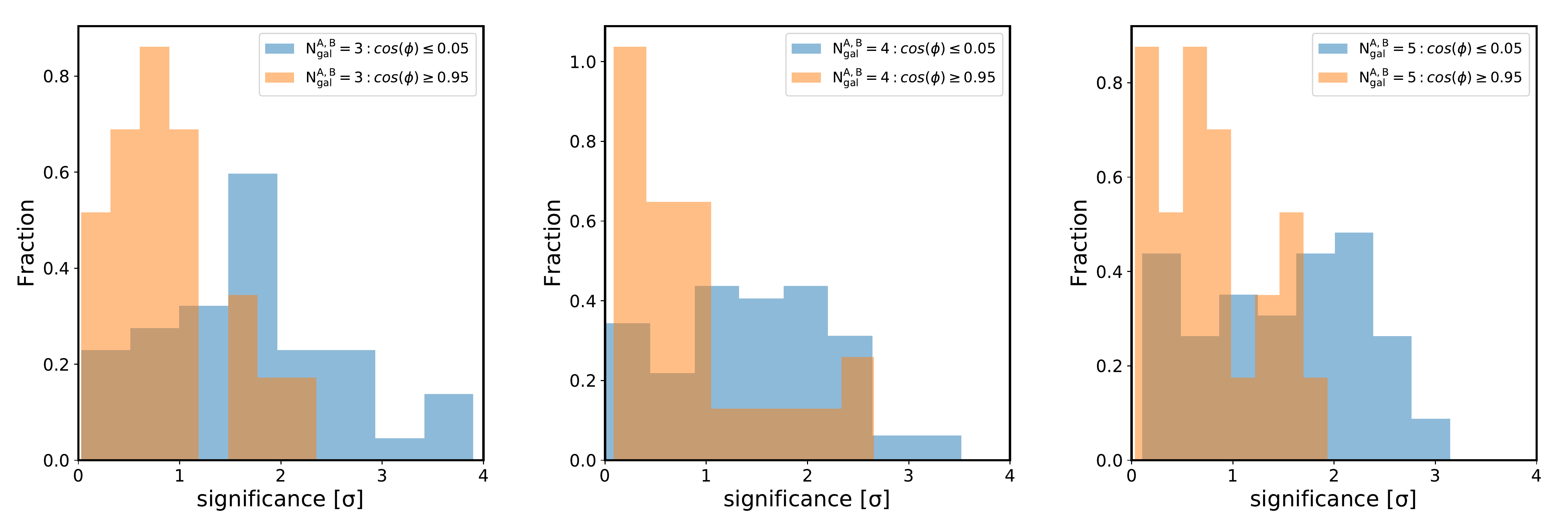}}
\caption{The distribution of significance (with respect to the random sample) for filaments that are inclined nearly parallel (blue) and nearly perpendicular (orange) to the line of sight. In the left, middle right panel we show this for thre choices of N$_{\rm gal} =3, ~4, ~5$. }
\label{fig:xxx}
\end{figure*}

In order to examine how the orientation of filaments changes the significance of the signal we wish to measure, we perform the following test. Let $\phi$ be the angle subtended between the filament spine and the line of sight. We compare filaments where $\cos\phi<0.05$ since these are nearly perpendicular to the line of sight with those where $\cos \phi > 0.95$ corresponding to filaments nearly parallel to the line of sight. These are shown by the blue and orange histograms (respectively) in Figure~\ref{fig:xxx}. Predictably we find that the filaments whose spines are closer to perpendicular to the line of sight have more significant rotation signals. Since the likelihood that a filament rotates does not depend on the orientation of the filament, we may ascribe the decrease in significance to projection effects.

Figure~\ref{fig:xxx} left, middle, and right panels consider the effect of changing the number of galaxies in each filament region A and B for 3, 4, and 5 galaxies respectively. Note that because each filament has two regions so this corresponds to a minimum of 6, 8 and 10 galaxies in the filament. We note that changing N$_{\rm gal}$ has little effect on the distribution of the estimated significance. Hence the choice of N$_{\rm gal} > 3 $ used throughout the paper is not biasing the result towards or away from significant detections.

\begin{figure*}[!ht]
    \centerline{\includegraphics[width=0.9\textwidth, trim=0.0cm 0.0cm 0.0cm 0.0cm,clip=true]{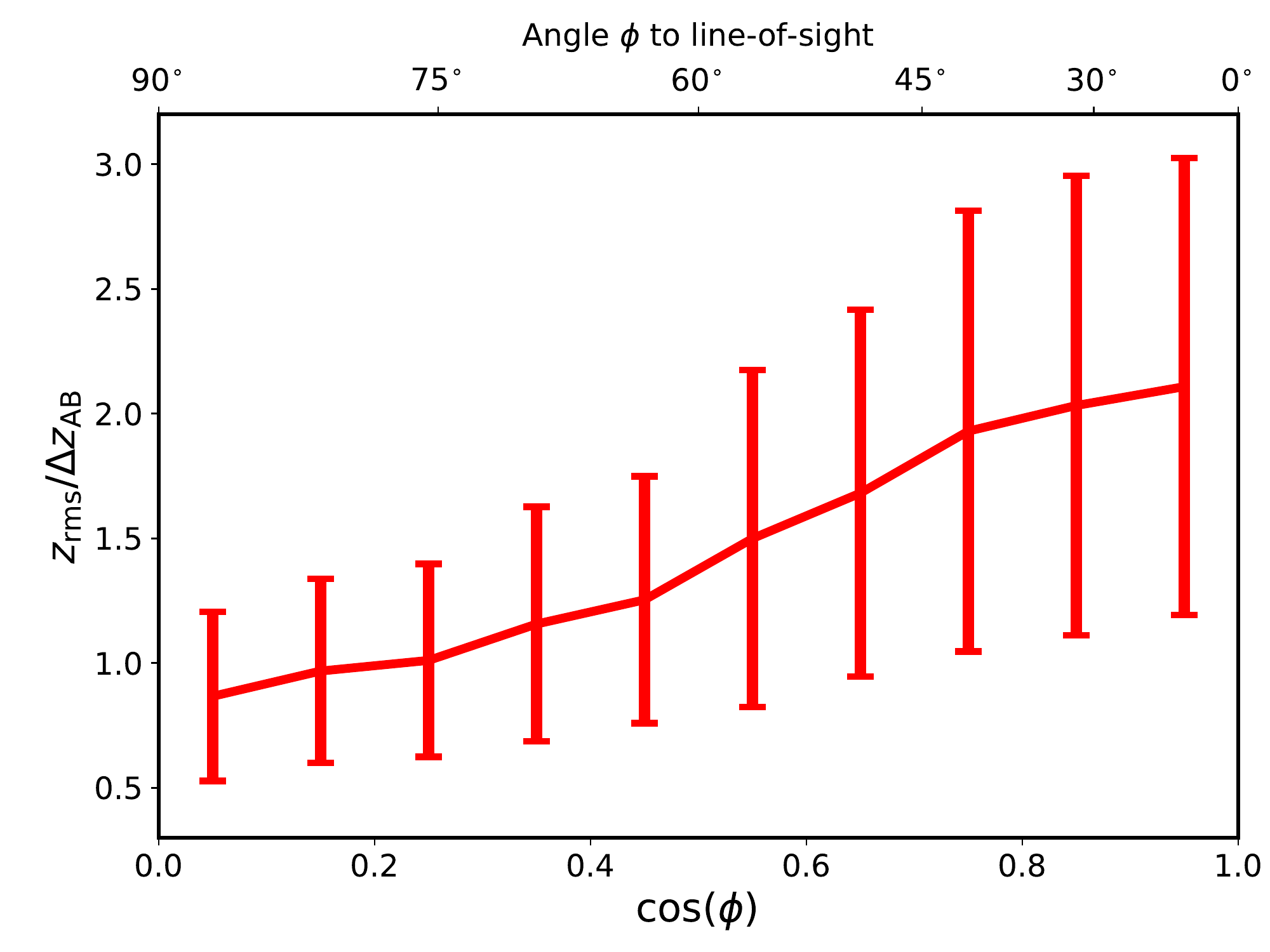}}
    \caption{{The median dynamical ``temperature'', $z_{\rm rms}/\Delta z_{\rm AB}$, of filaments as a function of the inclination angle $\phi$ between the filament spine and the line of sight. $z_{\rm rms}$ is the rms of galaxy redshift, and $\Delta z_{\rm AB}$ is the  mean redshift difference of galaxies on either side of the filament spine.} That this quantity increases as the filaments become parallel to the line of sight is consistent with galaxies moving predominantly around the filament spine. Error bars represent the standard deviation about the median.}
    \label{fig:zrms_dzab_angle}
\end{figure*}

\begin{addendum}

\item[Data Availability]  The galaxy data used in this paper is drawn from the publicly available SDSS DR12 (\url{https://www.sdss.org/dr12/}) and can be found here: \url{http://cosmodb.to.ee}. The Bisous filament catalogue can be found here: \url{http://cosmodb.to.ee}.

\item[Code availability] 
The codes used in this study are available from the corresponding authors upon reasonable request.

\item[Supplementary information]

\end{addendum}

\begin{flushleft}
\bf{\large References}
\end{flushleft}


\end{document}